\documentstyle[12pt]{article}

\font\titlefont=cmbx10 scaled \magstep4
\begin{document}

%\begin{flushright}
%\vspace*{1cm}
%Revised Version \\
%May 29, 2001
%\vspace*{1cm}
%\end{flushright}

\begin{center}
{\titlefont Stress Tensor Fluctuations and}
\vskip .1in
{\titlefont Passive Quantum Gravity}
\vskip .7in
L.H. Ford and Chun-Hsien Wu\\
\vskip .3in
Institute of Cosmology\\
Department of Physics and Astronomy\\
Tufts University\\
Medford, Massachusetts 02155\\
\end{center}

\vskip 1in
\baselineskip=14pt

\begin{abstract}
The quantum fluctuations of the stress tensor of a quantum field are discussed,
as are the resulting spacetime metric fluctuations. Passive quantum
gravity is an approximation in which gravity is not directly quantized,
but fluctuations of the spacetime geometry are driven by stress tensor
fluctuations. We discuss a decomposition of the stress tensor correlation
function into three parts, and consider the physical implications of each part.
The operational significance of metric fluctuations and the possible limits
of validity of semiclassical gravity are discussed.

\end{abstract}
\newpage

\section{Introduction}

The essential divide between classical gravity and the various quantum 
versions of gravity theory is crossed when the spacetime geometry
ceases to be fixed, but rather undergoes fluctuations. In a complete
theory of quantum gravity, one expects these fluctuations to arise
both from the quantum nature of gravity itself and from the quantum fluctuations
of matter fields which act as the source of gravity. The former are
``active'' (or spontaneous) fluctuations, whereas the latter are ``passive''
(or induced) fluctuations. A complete description of active spacetime
metric fluctuations would require a full quantum theory of gravity. 
However, it is possible to use linearized quantum gravity to describe
a variety of nontrivial phenomena, including quantum fluctuations of
the lightcone \cite{F95}. In this paper, we will focus upon the passive
metric fluctuations. Thus the gravitational field will not be quantized,
but nonetheless will undergo quantum fluctuations driven by matter fields.
This is what we mean by the phrase ``passive quantum gravity''.

The key to understanding passive quantum gravity is an analysis of the
fluctuations of the stress tensor of a quantized field, which will be
the principle topic of this paper. Most of our discussion will deal with
quantum fields on an approximately flat background. Fluctuations of the
quantum stress tensor were discussed in Refs.~\cite{F82,DF88,Kuo,WF99},
using an approach based on normal ordering, which will be discussed in more 
detail below. Other authors \cite{CH95,CCV,MV99} have discussed stress 
tensor fluctuations in the context of cosmology. 

In this paper, we will discuss a useful decomposition of the product
of stress tensor operators into three terms, a fully normal ordered term, a
cross term, and a vacuum term. The possible physical implications of each of
these terms will be considered in succession. In particular, we will
discuss how the cross term is responsible for the quantum fluctuations
of radiation pressure when a laser beam impinges upon a mirror, a potentially 
observable effect. We will also present some new results concerning the pure 
vacuum term. We calculate  both the
stress tensor correlation function and the resulting metric tensor 
correlation function in the Minkowski vacuum state, and show that the 
latter quantity can be expressed as total derivatives of a scalar
function. We then discuss the operation meaning of metric fluctuations
as the Brownian motion of test particles. We conclude with some remarks on 
the likely range of validity of semiclassical gravity in which metric
fluctuations are ignored.

\section{The Stress Tensor Correlation Function}

The basic object of interest is the quantum stress tensor operator,
$T_{\mu\nu}(x)$. However, this object is defined only after a renormalization.
That is, the formal expectation value of $T_{\mu\nu}(x)$ in any quantum
state is divergent. Fortunately, the divergence is a c-number, so the
renormalization is state-independent. The details of this procedure on a curved
background can be rather elaborate, and are discussed in many references
\cite{BD82}. For our purposes, it is sufficient to discuss the 
quantum stress tensor operator in Minkowski spacetime. In this case, the
c-number to be subtracted is simply the expectation value of $T_{\mu\nu}(x)$
in the Minkowski vacuum state, and the renormalized operator is the 
normal ordered operator:
\begin{equation}
:T_{\mu\nu}(x): = T_{\mu\nu}(x) - \langle T_{\mu\nu}(x) \rangle_0 \,.
\end{equation}
Here $\langle \;\rangle_0$ denotes the expectation value in the 
Minkowski vacuum state. 

In order to discuss stress tensor fluctuations, we must be able to define
the correlation function of a pair of renormalized stress tensor operators.
If we restrict ourselves to flat spacetime and normal ordered stress tensor 
operators, then we can define the correlation function as
\begin{equation}
C_{\mu\nu\rho\sigma}(x,x') = 
     \langle :T_{\mu\nu}(x): :T_{\rho\sigma}(x'): \rangle \, ,
\end{equation}
where the expectation value is understood to be taken in an arbitrary
quantum state. This correlation function can be decomposed into three
parts using Wick's theorem. The following identity can be established 
using this theorem:
\begin{eqnarray}
:\phi _{1}\phi _{2}::\phi _{3}\phi _{4}: & = & :\phi _{1}\phi 
_{2}\phi _{3}\phi _{4}:\nonumber \\
 &  & +:\phi _{1}\phi _{3}:\langle \phi _{2}\phi _{4}\rangle 
_{0}+:\phi _{1}\phi _{4}:\langle \phi _{2}\phi _{3}\rangle 
_{0}\nonumber \\
 &  & +:\phi _{2}\phi _{3}:\langle \phi _{1}\phi _{4}\rangle 
_{0}+:\phi _{2}\phi _{4}:\langle \phi _{1}\phi _{3}\rangle 
_{0}\nonumber \\
 &  & +\langle \phi _{1}\phi _{3}\rangle _{0}\langle \phi _{2}\phi 
_{4}\rangle _{0}+\langle \phi _{1}\phi _{4}\rangle _{0}\langle \phi 
_{2}\phi _{3}\rangle _{0}\, ,\label{eq:phiprod} 
\end{eqnarray}
where the \( \phi _{i} \) are free bosonic fields. 
In the remainder of this paper, we will assume that our stress tensor
operators are those of free bosonic fields, and hence can be expressed as
quadratic forms in the \( \phi _{i} \). 

We can now express the correlation function as
\begin{equation}
C^{\mu\nu\rho\sigma}(x,x') = C_{(N)}^{\mu\nu\rho\sigma}(x,x') +
C_{(cross)}^{\mu\nu\rho\sigma}(x,x') +C_{(V)}^{\mu\nu\rho\sigma}(x,x') \,.
\end{equation}
Here 
\begin{equation}
C_{(N)}^{\mu\nu\rho\sigma}(x,x') = 
\langle :T_{\mu\nu}(x) \,T_{\rho\sigma}(x'): \rangle
\end{equation}
is a fully normal ordered operator, and
\begin{equation}
C_{(V)}^{\mu\nu\rho\sigma}(x,x') = 
\langle :T_{\mu\nu}(x): :T_{\rho\sigma}(x'): \rangle_0
\end{equation}
is a pure vacuum term.  $C_{(cross)}^{\mu\nu\rho\sigma}(x,x')$ is a
cross term which is expressible as a sum of products of normal ordered
quadratic operators and vacuum expectation values of quadratic operators,
that is, products of the form of the middle four terms in 
Eq.~(\ref{eq:phiprod}).
The fully normal ordered term is state-dependent and finite in the coincidence
limit, $x' \rightarrow x$. The pure vacuum term is singular in this limit,
but is state-independent. However, the cross term is both state-dependent
and singular as $x' \rightarrow x$. Thus it is not possible to render
$C^{\mu\nu\rho\sigma}(x,x')$ finite by a state-independent subtraction, as 
it is $\langle T_{\mu\nu}(x) \rangle$. 

As we vary the quantum state to increase the mean energy density
$\langle :\rho: \rangle =  \\ \langle :T_{tt}: \rangle$, 
the fully normal ordered 
term will scale as $\langle :\rho : \rangle^2$, the cross term as 
$\langle :\rho: \rangle$, and the vacuum term does not change. Thus, in the 
limit of highly excited quantum states, the normal ordered term will
dominate. However, its contribution to the stress tensor fluctuations,
\begin{equation}
C^{\mu\nu\rho\sigma}(x,x') - \langle T_{\mu\nu}(x) \rangle
                             \langle T_{\rho\sigma}(x') \rangle
\end{equation}
need not grow any faster than that of the cross term.
The physical implications of each of the three terms will be discussd in
turn in the following sections.

\section{ The Fully Normal Ordered Term}

This term, as noted above, has the feature that it is finite in the
coincidence limit. If this were the only term present in the correlation 
function, then one could meaningfully discuss the fluctuations in local
stress tensor components, such as the energy density. This approach was used
by Kuo and Ford \cite{Kuo}, where only the fully normal ordered term
was retained, and a dimensionless measure of the local energy density
fluctuations was defined:
\begin{equation} 
\Delta = \frac{\langle :\rho^2: \rangle -\langle :\rho: \rangle^2}
              {\langle :\rho^2: \rangle} \,.  \label{eq:Delta}
 \end{equation}
In the case of a coherent state, $\Delta = 0$, so there are no fluctuations
in the local energy density by this measure. However, in non-classical
states, such as a squeezed vacuum state or a Casimir vacuum state, one can
have $\Delta$ of order unity. Similar results were found by Phillips and
Hu \cite{PH97} for the vacuum energy density in symmetric, curved spacetimes.

These results can be summarized by saying that 
$C_{(N)}^{\mu\nu\rho\sigma}(x,x')$ describes a fluctuating local energy density.
In the classical limit, these fluctuations vanish, but for non-classical
states, the fluctuations in the local enery density can be at least as large as
the mean energy density.

\section{ The Cross Term}

The simple picture of stress tensor fluctuations based upon 
$C_{(N)}^{\mu\nu\rho\sigma}(x,x')$ alone is not complete, in part because
of the existence of the cross term, $C_{(cross)}^{\mu\nu\rho\sigma}(x,x')$.
This term depends upon the quantum state, but is singular when
$x' \rightarrow x$. Thus it is not possible to define a local quantity
analogous to $\Delta$ which describes the effects of this term. However,
this does not mean that the cross term is devoid of physical meaning.
On the contrary, it is essential for understanding such phenomena as 
the quantum fluctuations of radiation pressure. 

\subsection{ Finiteness of Integrals of the Cross Term}

The singularity of the cross term need not be a concern if observable
quantities, which are space and time integrals, can be defined.
The cross term  goes as $(x-x')^{-4}$ as $x' \rightarrow x$. At first
sight, this is not an integrable singularity. However, it is in fact possible
to define the relevant integrals by an integration by parts procedure.
 The basic
idea can be illustrated as follows:
\begin{eqnarray}
\int_{-\infty}^\infty dt_1\, dt_2\, 
f(t_1) f(t_2) \frac{1}{(t_1-t_2)^4} 
&=& -\frac{1}{12} \int_{-\infty}^\infty dt_1\, dt_2\, f(t_1) f(t_2)
\frac{\partial^4}{\partial {t_1}^2 \partial {t_2}^2}\, \ln [(t_1-t_2)^2 \mu^2]
  \nonumber \\
&=& -\frac{1}{12} \int_{-\infty}^\infty dt_1\, dt_2\, 
\ddot{f}(t_1) \ddot{f}(t_2) \, \ln [(t_1-t_2)^2 \mu^2] \,, \label{eq:intpart}
\end{eqnarray}
where $\mu$ is an arbitrary constant. We have assumed that the function
$f(t)$ vanishes as $|t| \rightarrow \infty$, so the surface terms in the 
integration by parts vanish. The effect of this manipulation is to replace
the apparently non-integrable singularity in the first integral by a mild,
integrable singularity in the final integral. This trick has been employed
by various authors under the labels 
``generalized principal value integration'' \cite{Davies}
or ``differential regularization'' \cite{FJL}. 
Because the quantum state describes a distribution of energy which is limited 
in time, the normal ordered factors in 
$C_{(cross)}^{\mu\nu\rho\sigma}(x,x')$
vanish in both the past and the future, allowing the surface terms in the 
integrations to be dropped.

\subsection{ Quantum Fluctuations of Radiation Pressure}
\label{sec:rpfluct}

Classically, a beam of light falling on a mirror exerts a force and the 
force can be written as the integral of the Maxwell stress tensor. 
When we treat this problem quantum mechanically, then the force 
undergoes fluctuations. This is a necessary consequence of the fact
that physically realizable quantum states are not eigenstates of the
stress tensor operator. These radiation pressure fluctuations play an
important role in limiting the sensitivity of laser interferometer
detectors of gravitational radiation, as was first analyzed by 
Caves~\cite{Caves1,Caves2}. His approach was based on the statistical
fluctuations of photon numbers in a coherent state. Recently, we \cite{WF00}
have shown how this phenomenon can be understood in the context of the quantum
stress tensor. Here we will give a brief summary of this treatment.

Consider a mirror
of mass \( m \) which is oriented perpendicularly to the \( x 
\)-direction.
If the mirror is at rest at time \( t=0 \), then at time \( t=\tau  
\) its
velocity in the \( x \)-direction is given classically by 
\begin{equation}
\label{eq:v}
v=\frac{1}{m}\int _{0}^{\tau }dt\int _{A}da\, \, T_{xx}\, ,
\end{equation}
where \( T_{ij} \) is the Maxwell stress tensor, and \( \int _{A}da 
\) denotes
an integration over the surface of the mirror. Here we assume that 
there is
radiation present on one side of the mirror only. Otherwise, 
Eq.~(\ref{eq:v})
would involve a difference in \( T_{xx} \) across the mirror. When 
the radiation
field is quantized, \( T_{ij} \) is replaced by the normal ordered 
operator
\( :T_{ij}: \), and Eq.~(\ref{eq:v}) becomes a Langevin equation. The 
dispersion
in the mirror's velocity becomes
\begin{equation}
\label{eq:v2_{0}}
\langle \triangle v^{2}\rangle =\frac{1}{m^{2}}\int _{0}^{\tau }dt\, 
\int _{0}^{\tau }dt'\, \int _{A}da\, \int _{A}da'\, [\langle 
:T_{xx}(x)::T_{xx}(x'):\rangle -\langle 
:T_{xx}(x):\rangle \langle :T_{xx}(x'):\rangle ] \,.
\end{equation}

We now assume that the photons are in a single mode coherent state,
so that the fully normal ordered term gives no contribution. We are also
only interested in the changes in $\langle \triangle v^{2}\rangle$ due
to the radiation. Thus we can subtract off the Minkowski vacuum
contribution $\langle \triangle v^{2}\rangle_0$ and ignore the pure vacuum 
term. Now the entire contribution to the mirror's velocity fluctuations
comes from the cross term:
\begin{equation}
\label{eq:v2}
\langle \triangle v^{2}\rangle =\frac{1}{m^{2}}\int _{0}^{\tau 
}dt\int _{0}^{\tau }dt'\int _{A}da\int _{A}da'\langle 
T_{xx}(x)T_{xx}(x')\rangle _{cross}\, .
\end{equation}
The relevant component of the stress tensor is (Lorentz-Heaviside units 
are used here.) 
\begin{equation}
\label{eq:Txx}
T_{xx}=\frac{1}{2}(E_{y}^{2}+E_{z}^{2}+B_{y}^{2}+B_{z}^{2})\, .
\end{equation}
We now assume that a linearly polarized plane wave is normally 
incident and
is perfectly reflected by the mirror. Take the polarization vector to 
be in
the \( y \)-direction, so that \( E_{z}=B_{y}=0 \). At the location 
of the
mirror, \( E_{y}=0 \), and only \( B_{z} \) contributes to the stress 
tensor.
Thus, when we apply Eq.~(\ref{eq:phiprod}) to find \( \langle 
T_{xx}(x)T_{xx}(x')\rangle _{cross} \),
the only nonzero quadratic normal-ordered product will be \( \langle 
:B_{z}(x)B_{z}(x'):\rangle  \).
The result is 
\begin{equation}
\label{eq:Tcross2}
\langle T_{xx}(x)T_{xx}(x')\rangle _{cross}=\langle 
:B_{z}(x)B_{z}(x'):\rangle \langle B_{z}(x)B_{z}(x')\rangle _{0}\, .
\end{equation}
The vacuum magnetic field two-point function in the presence of a 
perfectly
reflecting plane at \( z=0 \) is given by 
\begin{equation}
\label{eq:2pt}
\langle B_{z}(t_{1},{\textbf {x}}_{1})B_{z}(t_{2},{\textbf 
{x}}_{2})\rangle _{0}=\langle B_{z}(t_{1},{\textbf 
{x}}_{1})B_{z}(t_{2},{\textbf {x}}_{2})\rangle _{E0}+\langle 
B_{z}(t_{1},{\textbf {x}}_{1})B_{z}(t_{2},{\textbf {x}}_{2})\rangle 
_{I0}\, .
\end{equation}
The first term is the two-point function for empty space, 
\begin{equation}
\label{eq:2pt_{e}mpty}
\langle B_{z}(t_{1},{\textbf {x}}_{1})B_{z}(t_{2},{\textbf 
{x}}_{2})\rangle _{E0}=\frac{(t_{1}-t_{2})^{2}+|{\textbf 
{x}}_{1}-{\textbf {x}}_{2}|^{2}-2(z_{1}-z_{2})^{2}}{\pi 
^{2}[(t_{1}-t_{2})^{2}-|{\textbf {x}}_{1}-{\textbf 
{x}}_{2}|^{2}]^{3}}\, .
\end{equation}
The second term is an image term 
\begin{equation}
\langle B_{z}(t_{1},{\textbf {x}}_{1})B_{z}(t_{2},{\textbf 
{x}}_{2})\rangle _{I0}=\langle B_{z}(t_{1},{\textbf 
{x}}_{1})B_{z}(t_{2},{\textbf {x}}_{2})\rangle _{E0}\biggl 
|_{z_{2}\rightarrow -z_{2}}\, .
\end{equation}
Both terms give equal contributions to the radiation pressure 
fluctuations on
a mirror located at \( z=0 \).

We can see that the integrand in Eq.~(\ref{eq:v2}) is singular
when the points $(t_{1},{\textbf {x}}_{1})$ and $(t_{2},{\textbf {x}}_{2})$
are lightlike separated from one another. However, this singularity
can be handled either by the integration by parts method of the previous
subsection, or equivalently by treating the integrals as containing
higher order poles. The result is (see Ref.~\cite{WF00} for details)
\begin{equation}
\label{eq:dv2_{s}t}
\langle \triangle v^{2}\rangle =4\, \frac{A\omega \rho }{m^{2}}\, 
\tau \, ,
\end{equation}
where $A$ is the illuminated area of the mirror, and $\rho$ is the mean
energy density in the laser beam.

As noted above, this result can be found from considerations of photon
number fluctuations. However, in an approach based upon the quantum
stress tensor, it arises solely from the cross term. Laser interferometer
detectors of gravity waves will eventually have to contend with radiation
pressure fluctuations as a noise source. At that point, it is reasonable to
expect that these fluctuations will be observed experimentally for 
the first time. Such
an observation would constitute experimental proof of the reality of
the cross term. 

It is of interest to note that if the quantum state is taken to be
a photon number eigenstate, rather than a coherent state, then 
the fully normal ordered term gives a non-zero contribution. However,
this contribution is such as to exactly cancel the contribution coming
from the cross term, leaving no radiation pressure fluctuations \cite{WF00}.

\section{The Pure Vacuum Term}

The piece of the stress tensor correlation function which is the most
difficult to interpret is the pure vacuum part, 
$C_{(V)}^{\mu\nu\rho\sigma}(x,x')$. This term is not only highly
divergent in the coincidence limit, but is always present. Any physical
effects which it produces would have to be very small so as not to have
already been observed. In this section, we will show that it can be written 
as a total derivative. 

\subsection{Explicit Form for the Electromagnetic Field}

The stress tensor of EM field is
\begin{equation}
\label{eq:stress}
T_{\mu \nu }=F_{\mu }\, ^{\rho }F_{\nu \rho }-
\frac{1}{4}g_{\mu \nu }F_{\alpha \beta }F^{\alpha \beta }\, ,
\end{equation}
where \( F_{\alpha \beta }=\partial _{\alpha }A_{\beta }-
\partial _{\beta }A_{\alpha } \). Expand the stress tensor in terms of 
the vector potential \( A_{\mu } \) to find
\begin{equation}
\label{eq:stress2}
T_{\mu \nu }=\partial _{\mu }A^{\rho }\partial _{\nu }A_{\rho }+
\partial ^{\rho }A_{\mu }\partial_{\rho }A_{\nu }-
\partial^{\rho }A_{\mu }\partial_{\nu }A_{\rho }-
\partial_{\mu }A^{\rho }\partial_{\rho }A_{\nu }-
\frac{1}{2}g_{\mu \nu }(\partial_{\alpha }A_{\beta }\partial^{\alpha }A^{\beta }
-\partial _{\alpha }A_{\beta }\partial ^{\beta }A^{\alpha })\,.
\end{equation}
In the Lorentz gauge, 
\begin{equation}
\langle A^{\mu}(x)A^{\nu}(x')\rangle _{0} = -g^{\mu \nu }\, D(x-x')\,,
\end{equation}
where
\begin{equation}
D = D(x-x') = \frac{1}{4 \pi^2 (x-x')^2} \label{eq:scalar_2pt}
\end{equation}
is the Hadamard (symmetric two-point) function for the massless scalar field.
We can see from Eq.~(\ref{eq:phiprod}) that
\begin{eqnarray}
\langle :A_{\mu}(x)A_{\nu}(x)::A_{\rho}(x')A_{\sigma}(x'):\rangle _{0}&=&
\langle A_{\mu}(x)A_{\rho}(x')\rangle_{0}
\langle A_{\nu}(x)A_{\sigma}(x')\rangle_{0}   \nonumber \\
&+& \langle A_{\mu}(x)A_{\sigma}(x')\rangle_{0}
\langle A_{\nu}(x)A_{\rho}(x')\rangle_{0} \, .
\end{eqnarray}
We can now combine these various relations to write, after some calculation,
an expression for the vacuum stress tensor correlation function
\begin{eqnarray}
C_{(V)}^{\mu\nu\sigma\lambda}(x,x') &=&
4\, (\partial_{\mu }\partial_{\nu }D)\, 
(\partial_{\sigma }\partial _{\lambda }D)\, 
+ 2\, g_{\mu \nu }\, (\partial_{\sigma }\partial_{\alpha }D)\, 
(\partial_{\lambda}\partial^{\alpha }D)\,
+ 2\, g_{\sigma \lambda}\, (\partial_{\mu }\partial_{\alpha }D)\, 
(\partial_{\nu}\partial^{\alpha }D)\,                 \nonumber \\
&-& 2\, g_{\mu \sigma }\, (\partial_{\nu }\partial_{\alpha }D)\, 
(\partial_{\lambda}\partial^{\alpha }D)\,
- 2\, g_{\nu \sigma }\, (\partial_{\mu }\partial_{\alpha }D)\, 
(\partial_{\lambda}\partial^{\alpha }D)\,              \nonumber \\
&-& 2\, g_{\nu \lambda }\, (\partial_{\mu }\partial_{\alpha }D)\, 
(\partial_{\sigma }\partial^{\alpha }D)\,
- 2\, g_{\mu \lambda }\, (\partial_{\nu }\partial_{\alpha }D)\, 
(\partial_{\sigma }\partial^{\alpha }D)\,            \nonumber \\
&+&(g_{\mu \sigma }g_{\nu \lambda }+g_{\nu \sigma }g_{\mu \lambda }-
 g_{\mu \nu }g_{\sigma \lambda })\,(\partial_{\rho }\partial_{\alpha }D)\, 
(\partial^{\rho}\partial^{\alpha }D) \,. \label{eq:CV}
\end{eqnarray}
A similar result for the case of the scalar field has been given by
Martin and Verdaguer  (see Eq.~3.42 of Ref. \cite{MV99}).

\subsection{The Metric Fluctuation Correlation Function}

We can now use our expression for the stress tensor correlation function
to find the correlation function for the passive metric fluctuations
induced by vacuum fluctuations of the electromagnetic field. Let $h_{\mu \nu }$
be a classical metric perturbation due to the stress tensor $T_{\mu \nu }$.
Define \( \bar{h}_{\mu \nu }=h_{\mu \nu }-\frac{1}{2}\eta _{\mu \nu }h \)
and impose the harmonic gauge condition, 
(\( \partial _{\nu }\bar{h}^{\mu \nu }=0 )\). Then
\begin{equation}
\Box \bar{h}_{\mu \nu }=-16\pi T_{\mu \nu }\,
\end{equation}
in units in which $G =1$, where $G$ is Newton's constant. Let
\( G_{r}(x-x') \) be the retarded Green function which satisfies
\begin{equation}
\label{eq:delta}
\Box G_{r}(x-x')=\delta (x-x')\, .
\end{equation}
If there is no incoming gravitational radiation, \( \bar{h}_{\mu \nu }(x) \) is given
by
\begin{equation}
\bar{h}_{\mu \nu }(x)=-16\pi \int d^{4}x_{1}G_{r}(x-x_{1})T_{\mu \nu }(x_{1})
\, .
\end{equation}

Now let \( T_{\mu \nu } \) be the normal-ordered stress operator for the 
quantized electromagnetic field. Because here $T^\mu_\mu = 0$, we have
$\bar{h}_{\mu \nu } = {h}_{\mu \nu }$.
The metric fluctuation correlation function
is now
\begin{equation}
\label{eq:h-2point}
\langle h^{\mu \nu }(x)h^{\rho \sigma }(x')\rangle =
(16\pi )^{2}\int d^{4}x_{1}d^{4}x_{2}\, G_{r}(x-x_{1})G_{r}(x'-x_{2})
\, C_{(V)}^{\mu\nu\rho\sigma}(x_1,x_2) \, .
\end{equation}
We use Eqs.~(\ref{eq:scalar_2pt}) and (\ref{eq:CV}) in the above expression.
The result may be written in terms of derivatives of the quantity
\begin{equation}
S = \ln^{2}[\mu^{2}(x-x')^{2}]\, , 
\end{equation}
where $\mu$ is an arbitrary constant, using results such as
\begin{equation}
\label{eq:square}
\Box^2 S= -\frac{32}{[(x-x')^2]^2}\, . 
\end{equation}
Finally we   perform a set of integrations by parts and   assume
that the surface terms can be ignored. (This assumption needs to be
examined more carefully, and is a current topic of investigation.)
More details of the calculation will be given in a later paper.
The final result for the metric correlation function is
\begin{eqnarray}
\langle h_{\mu \nu }(x)h_{\sigma \lambda}(x')\rangle  & = & 
-\frac{1}{60\, \pi^2}\Bigl[
 4\, \partial_{\mu }\partial_{\nu }\partial_{\sigma }\partial_{\lambda }\, S
+ 2 \,(g_{\mu \nu }\, \partial_{\sigma }\partial_{\lambda }
   + g_{\sigma \lambda}\, \partial_{\mu }\partial_{\nu} )\, \Box S \nonumber \\
&-& 3\, (g_{\mu \sigma }\, \partial_{\nu }\partial_{\lambda} +
     g_{\mu \lambda }\, \partial_{\nu }\partial_{\sigma} +
     g_{\nu \sigma }\, \partial_{\mu }\partial_{\lambda} +
     g_{\nu \lambda }\, \partial_{\mu }\partial_{\sigma})\, \Box S \nonumber \\
&+& 3\, (g_{\mu \sigma }g_{\nu \lambda }+
                              g_{\nu \sigma }g_{\mu \lambda })\,\Box^2 S
- 2\, g_{\mu \nu }g_{\sigma \lambda }\,\Box^2 S \Bigr]\,.    \label{eq:h-h-2p} 
\end{eqnarray}
This is a remarkably simple result. It is of 
special interest to note 
that the metric fluctuation correlation function is expressible as
the total derivative of a scalar.

\section{Operational Meaning of Metric Fluctuations}

Fluctuations of the spacetime metric ultimately must be recorded by
test particles or waves propagating in the fluctuating geometry.
Let us first consider the use of a classical point test particle.
In classical relativity, such a test particle moves on a geodesic
in a fixed classical metric and can serve as giving operational meaning
to the spacetime geometry. If we now allow the metric to fluctuate,
the geodesic equation becomes a Langevin equation and the test particles
undergo Brownian motion \cite{Kuo}. We can express this Langevin equation
as 
\begin{equation}
\frac{d u^\mu}{d \tau} = - \Gamma^\mu_{\alpha\beta}u^\alpha u^\beta
                         - \gamma^\mu_{\alpha\beta}u^\alpha u^\beta \,, 
    \label{eq:Langevin}
\end{equation}
where $u^\mu$ is the particle's four-velocity, 
$\Gamma^\mu_{\alpha\beta}$ is the connection due to the
mean metric, and $\gamma^\mu_{\alpha\beta}$ is the linear
correction to the connection due to the fluctuations. Thus
\begin{equation}
\langle \gamma^\mu_{\alpha\beta} \rangle = 0 \,.
\end{equation}
We may integrate this equation, and then calculate mean squared variations
in the  four-velocity in terms of the metric fluctuation correlation
function, $\langle h^{\mu \nu }(x)h^{\rho \sigma }(x')\rangle$. Note that
this correlation function is given in passive quantum gravity by the
generalization of Eq.~(\ref{eq:h-2point}):
\begin{equation}
\label{eq:h-2point_gen}
\langle \bar{h}^{\mu \nu }(x) \bar{h}^{\rho \sigma }(x')\rangle =
(16\pi )^{2}\int d^{4}x_{1}d^{4}x_{2}\, G_{r}(x-x_{1})G_{r}(x-x_{2})
\, C^{\mu\nu\rho\sigma}(x_1,x_2) \, ,
\end{equation} 
where now the full stress tensor correlation function 
$C^{\mu\nu\rho\sigma}(x_1,x_2)$ appears. 
This procedure allows us to calculate such quantities as the mean
angular deflection or the mean time delay or advance due the the fluctuating
metric. 

Instead of a point particle, one might use classical waves as the probes
of the fluctuating geometry \cite{HS98}. In this case, one could write down a 
correction to a solution of a wave equation due to linearized metric
perturbations, which plays a role analogous to the $\gamma^\mu_{\alpha\beta}$
term in Eq.~(\ref{eq:Langevin}). This term will produce fluctuations
in the wave intensity at a given observation point. There is a need
for more detailed model calculations to better understand both the test
particle and the wave approaches to probing a fluctuating geometry.

\section{Validity of the Semiclassical Theory of Gravity}

The semiclassical theory of gravity assumes a fixed spacetime metric
satisfying the semiclassical Einstein equation
\begin{equation}
G_{\mu\nu} = 8 \pi \langle T_{\mu\nu} \rangle \,. \label{eq:semi}
\end{equation}
This equation is clearly an approximation which must fail at some point. 
First, it does not include any effects of the quantization of gravity
itself, the active metric fluctuations. However, even if we restrict 
ourselves to situations where only the quantum effects of matter fields
are included, Eq.~(\ref{eq:semi}) must fail when the passive metric
fluctuations become too large. The question is, how large is too large?

Kuo and Ford \cite{Kuo} suggested that a possible criterion could be based
upon the quantity $\Delta$ defined in Eq.~(\ref{eq:Delta}). If $\Delta \ll 1$,
then the fractional fluctuations in the local energy density, as
measured by $C_{(N)}^{\mu\nu\rho\sigma}(x,x')$ are small, and one
expects the resulting metric fluctuations also to be small. However,
if $\Delta$ is not small, then there are large local energy density
fluctuations. Kuo and Ford took 
\begin{equation}
\Delta \ll 1     \label{eq:KF}
\end{equation}
 as a necessary condition 
for the validity of the semiclassical theory. This criterion has been
criticized by Phillips and Hu \cite{PH00} as being too strong. The
latter authors calculate a quantity analogous to $\Delta$, but involving
smeared fields in the Minkowski vacuum state. They find that this quantity
is of order one. Because one expects the semiclassical theory to be valid
in Minkowski spacetime, Phillips and Hu conclude that  Eq.~(\ref{eq:KF})
is not a reliable criterion.

We wish to give an  assessment both of the Kuo-Ford criterion and
of Phillips and Hu criticism of it. First, it now seems that the 
Kuo-Ford criterion is at best incomplete because it does not address the 
effects of the cross term. The radiation pressure fluctuations studied
in Sect.~\ref{sec:rpfluct} show that this term has physical reality and 
must contribute to quantum metric fluctuations. The extent of its contribution
is not yet clear. However, in some model radiation pressure calculations
for thermal states \cite{WF99} and in the Casimir effect \cite{WKF01},
the cross term gives a larger contribution than does the fully normal
ordered term. Furthermore, the real effect of both terms on metric 
fluctuations is measured by integrals along the worldlines of test particles
rather than by local quantities. 

However, the analysis of Phillips and Hu is open to the critcism that the
quantites which they define are not directly observable. The type of
averaging which is involved in a measurement of a fluctuating spacetime
by test particles is more of the form of that in Eq.~(\ref{eq:h-2point_gen})
than of smearing field operators themselves. This leads us to the question of
whether the pure vacuum term can have observable effects in Minkowski spacetime.
The metric fluctuation correlation function given
in Eq.~(\ref{eq:h-h-2p}) is the total derivative of a scalar. This suggest that
when one uses it to calculate the Brownian motion of test particles or the 
fluctuation in amplitude of a wave, the result can be cast into the form 
of a surface term by an integration by parts. However, surface terms can 
be made to vanish when quantites such as the wave amplitude are switched
on in the past and off in the future. This is by no means a rigorous argument,
but rather a heuristic suggestion that the pure vacuum term may not
produce observable effects. This suggestion needs to be tested by more
detailed analysis. If it is correct, then Phillips and Hu
criticism of the Kuo-Ford criterion is muted. 

This would still not necessarily mean that the Kuo-Ford criterion is a good 
measure of the effects of metric fluctuations. As noted above, it ignores
the effects of the cross term. More generally, it now seems that
any criterion for the validity of the semiclassical theory must be
a non-local one. That is, it should involve integrals upon the worldlines
of test particles. It is possible that one can have situations where
there are large fluctuations on short time or distance scales, but which 
average out when measurements on longer scales are made. 

If the vacuum term is indeed unobservable, then one must study in detail
 the combined effects of the normal ordered and the cross
term on the Brownian motion of test particles. This also remains to be
done. In the end, the validity of the semiclassical theory will probably
depend on the question which one wishes to answer. If one is interested
only in quantites averaged over scales large compared to the intrinsic
scales defined by the quantum state, then the semiclassical theory may
well give an accurate answer. However, if one poses a question about
behavior on shorter scales, the fluctuations are more likely to be
important. A useful analogy is the fluctuating mirror discussed in
Sect.~\ref{sec:rpfluct}. If one is only interested in the average motion
of the mirror, then Newton's second law with the mean force is adequate.
However, if one needs to know the position of the mirror to high
accuracy, as in a sensitive interferometer, then the force fluctuations
cannot be ignored.

In summary, it seems likely that the validity of the semiclassical
approximation will depend upon several factors. First, it depends upon
what question one is asking. This determines the level at which one decides 
that the effects of fluctuations around a mean geometry are negligible.
Second, it can depend upon the choice of quantum state. We have seen that
the fluctuations of the normal ordered term are minimized in a coherent state,
but can be large in other states. Similarly, radiation pressure fluctuations 
are minimized in a photon number eigenstate, but can be significant in other
states. Finally, the magnitude of fluctuation effects depends upon time
and length scales, which can in turn depend upon the quantum state.
 Measurements which average over larger scales have a greater
tendency to average out the effects of fluctuations than do those made on very 
short scales.

\vspace{0.5cm}

 {\bf Acknowledgements:} We would like to thank B.L. Hu, C.I. Kuo, 
N.G. Phillips, and E. Verdaguer for stimulating discussions. 
This work was supported in part by the National
Science Foundation under Grant PHY-9800965.

\end{document}